\documentclass[aps,prl,twocolumn,showpacs, superscriptaddress]{revtex4}
\usepackage{graphicx,amssymb,amsfonts,amsmath}

\begin{document}

\title{Current-driven quantum criticality in itinerant electron ferromagnets
}
\author{Aditi Mitra}
\affiliation{Department of Physics, New York University, 4
Washington Place, New York, NY 10003 USA
}
\author{Andrew J. Millis}
\affiliation{Department of Physics, Columbia University, 538 W.
120th Street, New York, NY 10027 USA
}
\date{\today}


\begin{abstract}
We determine the effect of an in-plane
current flow on the critical properties of a 2d itinerant electron system near a ferromagnetic
-paramagnetic quantum critical point. 
We study a model in which 
a nonequilibrium steady state is established as a result of exchange 
of particles and energy with an underlying substrate. 
The current ${\vec j}$ gives rise not only to an effective temperature
equal to the voltage drop over a distance of order the mean free path,
but also  to symmetry-breaking
terms of the form $ {\vec j}\cdot \vec{\nabla}$ in the effective action.
The effect of the symmetry breaking on the fluctuational and critical properties is 
found to be small although 
(in agreement with previous results)
if rotational degrees of freedom are important, the current
can make the classically ordered state dynamically unstable.

\end{abstract}

\pacs{73.23.-b,05.30.-d,71.10.-w,71.38.-k}

\maketitle

One of the most intriguing  issues
in modern condensed matter physics is quantum criticality,
the complex phenomena associated with phase changes driven by variation of
Hamiltonian parameters such as pressure, magnetic field, or chemical composition 
in the limit as temperature $T \rightarrow 0$~\cite{Hertz76,Millis93,Sondhi97,Sachdev99}. Quantum phase transitions
differ from classical phase transitions because the 
non-commutativity  of position and momentum in quantum mechanics implies that spatial and temporal  fluctuations 
are coupled  at a quantum critical point.
While quantum phase transitions in equilibrium systems have been the 
subject of extensive study, phase transitions caused by nonequilibrium drives
(for example an imposed current flow) have been much less extensively studied.
Scaling theories and calculations have shown that one important effect of a
departure from equilibrium  is to produce decoherence effects which may be
modeled as an effective temperature~\cite{Phillips04,Green06,Green05,Mitra06,TakeiKim07}. 
However, departures from equilibrium may lead to other effects \cite{Feldman05}.
In particular,    
a current drive breaks inversion symmetry, and it is well known that symmetry
plays a crucial role in critical phenomena.
In this paper we investigate the effect of a current drive on the critical properties
of a low dimensional system near a ferromagnetic quantum critical point. We present a derivation
from microscopics which reveals the inversion-symmetry-breaking terms, and we analyze their consequences.

\begin{figure}
\includegraphics[totalheight=3cm,width=6cm]{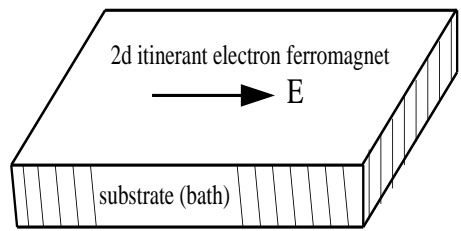}
\caption{A 2d itinerant electron ferromagnet driven out of equilibrium by application of an in-plane
electric field. A steady state is reached via coupling to an underlying substrate.
}
\label{schem}
\end{figure}

The geometry that will be studied (shown 
schematically in Fig~\ref{schem}) is a very thin layer laid on top
of a substrate with which the layer can exchange particles as well as energy.
We take the layer to be a metal which is tuned to be near a ferromagnetic-paramagnetic 
quantum critical point.
Departures from equilibrium and breaking of inversion symmetry are provided by current, 
$j$, flowing along the layer, while 
the coupling to the substrate allows the system to reach a nonequilibrium steady state. 
An earlier paper~\cite{Mitra06} studied a similar system but with
nonequilibrium provided by current flow across the layer, so the issue of inversion symmetry breaking did not arise. 

We now outline the relevant scales and summarize our major results. 
An electron in the layer will escape into the substrate after a typical time $\tau_{sc}$ determined by
the strength of the coupling to the substrate
and taken to be long compared to typical electronic times such as the inverse 
of the Fermi energy $E_F$: $E_F\tau_{sc}>>1$.
The escape time implies a length $l_{sc}=v_F\tau_{sc}$ with $v_F$
the Fermi velocity in the layer. A current $j$ flowing in the layer implies an electric field $E=\rho j$. 
The departure from equilibrium produces  decoherence whose consequences may be approximately
parametrized by  an effective temperature $T_{eff}$~\cite{Green05,Mitra06} which 
in the present problem
we find to be given  by the voltage drop over an escape length; 
\begin{equation}
T_{eff}=eEl_{sc}
\label{Teffdef}
\end{equation} 
We present here a theory valid in the limit $T_{eff}\tau_{sc}<<1$. 

The new feature of the longitudinal current is a breaking of inversion symmetry. 
The magnitude of the symmetry breaking effect in the spin sector turns out
to be related to the spin current $j_M=j_\uparrow-j_\downarrow$ induced by an applied
electric field. In the model we study we find that for a  
magnetization density $M$ the spin current is $j_M=v_D M$ with magnetization
drift velocity $v_D$ proportional both to the applied electric field~\cite{BJZ98,Macd04} and to 
deviations from particle-hole symmetry. 
The importance of particle-hole symmetry may be seen in a simple 
Boltzmann transport picture: the current $j_\sigma$
of electrons of spin $\sigma$ is ${\vec j}_\sigma=\int (dk) \vec{v}(k) n(\varepsilon_k-\mu_\sigma;E)$
with $n$ a distribution function depending on the applied electric field, the electron energy, and
the spin-dependent chemical potential. In a particle-hole symmetric situation, $v(k)$ is independent of $k$
and $\varepsilon_k=v|k|$, so a simple variable shift implies that $j_\uparrow=j_\downarrow$ implying $j_M=0$.
In order of magnitude $v_D k_F \sim T_{eff}<<E_F$, so that the effects of inversion symmetry breaking turn out to be small.

We write a Keldysh path integral~\cite{Keldysh63,Kamenev04},
make a saddle point approximation followed by 
an expansion of the Keldysh action in fluctuations about the 
saddle point~\cite{Mitra05a,Mitra06}. The resulting  
nonequilibrium Ginzburg-Landau theory is then 
treated by
renormalization group~\cite{Mitra06,Mitra07}.  
The Hamiltonian of the system 
is
\begin{equation}
H = H_{bath} + H_{layer} + H_{layer-bath} \label{hm}\\
\end{equation}
where $H_{layer}$
is the interacting electron layer whose critical properties we
are interested in,
$H_{bath}$ describes the underlying substrate which will be assumed to be
an electron gas which is always in thermodynamic equilibrium,  while $H_{layer-bath}$
couples the two.  
Thus, 
\begin{eqnarray}
H_{layer} &=& \sum_{\vec{p}_{\perp}\sigma}
\epsilon_{\vec{p}_{\perp} - \frac{e}{\hbar c}\vec{A}}
d^{\dagger}_{\vec{p}_{\perp}\sigma} d_{\vec{p}_{\perp}\sigma} +
U\sum_{i}
n_{i\uparrow} n_{i\downarrow} \label{hsys}\\
H_{layer-bath} &=& \sum_{\vec{p}_{\perp},\sigma}
\left(t
c^{\dagger}_{bath,\sigma}
d_{\vec{p}_{\perp}\sigma} + h.c.\right) \label{hcoup}
\end{eqnarray}
with $\sigma$ the spin label, $p_{\perp}$ the momentum within the layer which is 
a good quantum number, and $\vec{A} = - c\vec{E} t$. 
The only relevant features of $H_{bath}$ are its density of states, $N_{bath}$,
which implies that $\tau_{sc}=(2\pi N_{bath}t^2)^{-1}$ and its resistivity, which we take to be very high relative to that of the layer so we may couple
the electric field only to the layer electrons.
 
The interaction term is treated 
via the usual Hubbard-Stratonovich transformation, which involves
introducing Hubbard-Stratonovich fields $m_{i\pm}$ 
in the time-ordered 
(denoted by $-$) and anti-time ordered (denoted by
$+$) Keldysh axes~\cite{Mitra06},
choosing a quantization axis at each point in space and time~\cite{Schulz90} 
and introducing appropriate  spin rotation matrices ${\bf R_{\pm}}$ .  To simplify the notation
we do not write these explicitly except where needed.  

After integrating out the electronic degrees of freedom
in the bath 
and layer we obtain the following Keldysh action for 
the classical ($m_{cl}= \frac{m_ - + m_+}{2}$) and quantum components ($m_q = \frac{m_- - m_+}{2}$)
of the 
auxiliary fields
\begin{eqnarray}
Z_K = \int {\cal D} \left[m_q m_{cl} {\cal D}\left[\Omega\right] \right] e^{-i \int d^d x dt \, 2 U m_q m_{cl}} \nonumber \\ 
\times e^{Tr\ln \left[i\left( G^{-1}_{\sigma} - U \sigma \begin{pmatrix} m_{cl} & m_q \\ m_q & m_{cl} 
\end{pmatrix}\right)\right]} 
\label{SK1}
\end{eqnarray}
where 
the Keldysh Greens function for the layer electrons 
\begin{equation}
G = \begin{pmatrix}G^R & G^K\\0 & G^A \end{pmatrix}.
\end{equation}

The effect of the nonequilibrium drive is expressed via the E-field dependence of $G$, which
we now discuss. 
The solution of the Dyson equation for the retarded/advanced
component of $G$ is 
\begin{equation}G^{R/A}(\vec{p},t_1,t_2) = \mp i \theta\left(\pm(t_1-t_2)\right) e^{-i \int^{t_1}_{t_2} dx \epsilon(\vec{p} + e \vec{E}x)} e^{-\frac{|t_1 -t_2|}{2\tau_{sc}}}
\label{gr}
\end{equation} 
Introducing the time difference $\tau = t_1-t_2$
and  canonical momentum $\vec{k}=\vec{p} + e \vec{E} T$, ($T = \frac{t_1 + t_2}{2}$)
we find that up to terms of order $ (eE)^2(\tau_{sc}^3\partial^2\varepsilon_k/(\partial k^2))$ in the exponential,
the retarded/advanced components of the layer Green function $G$ take
the equilibrium form 
\begin{equation}
G^{R/A}(p,t_1,t_2;E)\rightarrow G^{R/A}(\vec{k}, \tau) = \mp i\theta(\pm\tau) e^{-i \epsilon_k \tau} 
e^{-\frac{|\tau|}{2\tau_{sc}}}
\label{GR}
\end{equation}
By introducing the lattice constant $a$ and noting that  the fermi energy $E_F \sim v_F/ a$,
we may 
estimate the magnitude of  the neglected terms as  $\frac{\partial^2\epsilon/a^2\partial k^2}{E_F}\frac{ (T_{eff}\tau_{sc})^2}{E_F\tau_{sc}}$
which is much less than unity in the limit of relevance here. 

The Keldysh Green's function
$G^K = G^R\cdot f_K - f_K \cdot G^A$ where
the distribution function $f_K= 1-2 f$ obeys
\begin{eqnarray}
&&i \frac{\partial f_K}{\partial t_1} + i \frac{\partial f_K}{\partial t_2}-\epsilon_{p-\frac{e}{\hbar c}A(t_1)}f_K 
+ \epsilon_{p-\frac{e}{\hbar c}A(t_2)} f^K \nonumber \\
&&- \Sigma^R \cdot f_K + f_K \cdot \Sigma^A 
+ \Sigma^K=0
\label{eomf}
\end{eqnarray}
$\Sigma^R - \Sigma^A =-\frac{i}{\tau_{sc}}$ and $\Sigma^K = (\Sigma^R - \Sigma^A)(1-2g)$,
$g$ being the distribution function of the substrate. 
Fourier transforming Eq.~\ref{eomf} with respect to the relative time $\tau=t_1-t_2$, changing variables
to the canonical momentum $\vec{k} = \vec{p}+ e \vec{E}T$ and expanding in $E$ 
one finds that  the distribution function at 
steady state obeys,
\begin{eqnarray}
&&e\vec{E}\cdot \frac{\partial f}{\partial \vec{k}} + \frac{\partial f}{\partial \omega} \left(e\vec{E}\cdot\frac{\partial \epsilon_k}{\partial
\vec{k}}\right)
 + \frac{1}{24}\frac{\partial^3 f}{\partial \omega^3}\left(e\vec{E}\cdot \frac{\partial}{\partial \vec{k}}\right)^3
\epsilon_k 
\ldots\nonumber \\
&&=\frac{1}{\tau_{sc}}\left[-f + g \right]
\label{eomf2}
\end{eqnarray}
The usual quasiclassical arguments~\cite{Rammer86} imply that the first term in Eq \ref{eomf2} is negligible
while in the  the weak field limit  the third term may be dropped. 
With these simplifications we find
$f=f^s + f^a$ where
\begin{eqnarray}
f^s_{k,x}&=& \theta(-x) + \frac{sign(x)}{2}
e^{- \frac{|x|}{\sqrt{\left(e\vec{E}\cdot \vec{v}_k \tau_{sc}\right)^2}}}
\label{fssolph}\\
f^a_{k,x} &=&\frac{\left(e \vec{E} \cdot \vec{v}_k \tau_{sc}\right)}{2\sqrt{\left(e\vec{E}\cdot \vec{v}_k \tau_{sc} \right)^2}} 
e^{- \frac{|x|}{\sqrt{\left(e\vec{E}\cdot 
\vec{v}_k \tau_{sc}\right)^2}}} 
\label{fasolph}
\end{eqnarray}
where $x = \omega - \mu$ and $v_k=\partial \varepsilon_k/\partial k$. Substitution of Eqs
\ref{fssolph}, \ref{fasolph} into Eq \ref{eomf2} then shows that the neglect of the third term
in Eq \ref{eomf2} is justified 
at weak layer-substrate coupling 
$( \frac{\partial^3 \epsilon_k}{E_Fa^3 \partial k^3}\ll \left(E_F \tau_{sc}\right)^2)$  while the first term
is negligible in the weak field limit $T_{eff} \ll E_F^2/(\frac{\partial^2 \epsilon_k}{a^2\partial k^2})$.

The quantum critical analysis proceeds ~\cite{Kamenev04,Mitra06} by finding a saddle point with
$m_q=0$ and an $m_{cl}$
that obeys 
\begin{equation}
2iU m_{cl} = -U Tr \left[\left(G^{-1}_{\sigma} - \sigma U m_{cl} \right)^{-1} \sigma \tau_x\right]
\label{saddle}
\end{equation} 
where $\tau$ are Pauli matrices in Keldysh space,
and then expanding about the saddle point and 
using renormalization group methods to treat the resulting action. 
In the disordered
phase the saddle point is $m_{cl}=m_q=0$ and by expanding the
Trln in Eq.~\ref{SK1} we obtain
\begin{eqnarray}
Z_K = 
\int {\cal D} \left[m_q m_{cl} \right] e^{-i \int d^d x dt \, 2 U m_q m_{cl}} e^{S^2_K + S^3_K+S^4_K + \ldots}
\label{SK2}
\end{eqnarray}
where 
\begin{eqnarray}
S_K^2 = -iU\int \frac{d^dq}{(2\pi)^d} \frac{d\Omega}{2\pi}
\left[m_q^* m_{cl} \Pi^R + c.c. \right] 
+ m_q m_q^* \left[\Pi_K\right] 
 \label{SKeff2} 
\end{eqnarray}
involves polarization bubbles $\Pi_{R/A/K}$ computed in terms of the $G$. For the Ising 
case $S_K^3=0$,
while $S_K^4 = \sum_{i=1\ldots 4} u_i m_q^i m_{cl}^{4-i}$.
We have used Eqs ~\ref{GR}, ~\ref{fssolph} and ~\ref{fasolph} to compute the $\Pi$ and $u_i$~\cite{Mitra06}.
Expanding
in $q$, $\Omega$, distance from criticality $\delta$ and  $ T_{eff}\tau_{sc}$ we find
\begin{equation}
U\Pi^R(q,\Omega) = \delta + A q^2 -i\tau_{sc}\left (\Omega-{\vec v}_D\cdot {\vec q}\right)  \label{piReff}
\end{equation}
The  
$i(\Omega-\vec{v}_D \cdot \vec{q})\tau_{sc}$ factor expresses the non-conserved 
dynamics induced by the 
escape of electrons from the layer into the substrate; the form of this
term shows that the fluctuations drift at the magnetization drift velocity $v_D$
which in the approximations considered here is
\begin{equation}
{\vec v}_D= e\vec{E} \tau_{sc} \frac{1}{N_0} \int \frac{d^d k}{(2\pi)^d} \delta(\epsilon_k -\mu) 
\frac{\partial^2 \epsilon_k}{\partial k^2}
\label{vmdef}
\end{equation}
where $N_0 = \int \frac{d^dk}{(2\pi)^d} \delta(\epsilon_k - \mu)$.
The drift term
was missed
in previous work~\cite{Green06}. To estimate its order of magnitude
we note that Eq~\ref{vmdef} involves the scale $T_{eff}$ and bare quantities. 
Because one may regard the dimension of velocity as $[Energy \times Length]$
and the only available length is the fermi wave vector, we estimate
$v_D = T_{eff}/k_F$.

The Keldysh polarization bubble contains information about decoherence and noise. 
In contrast to  $\Pi^{R/A}$, $\Pi^K$ is found to depend on $\Omega$ only
because the noise arises from coupling to the substrate and does not
drift in the presence of a current in the layer.  We find in 2d  

\begin{eqnarray}
\Pi^K_{2d}(\Omega)& &= -2i\tau_{sc} \label{PiK2d}  \\
&& \left[ |\Omega| + T_{eff}\int_{-\pi}^\pi \frac{d\phi}{2\pi} |cos\phi| e^{-\frac{|\Omega|}{(T_{eff}|cos\phi|)}}\right]
\nonumber
\end{eqnarray}
and for a 1d system,
\begin{eqnarray}
\Pi^K_{1d}(\Omega)& &= -2i\tau_{sc}
\left[ |\Omega| + T_{eff} e^{-\frac{|\Omega|}{T_{eff}}}\right]  
\label{PiK1d} 
\end{eqnarray}

Eqs~\ref{PiK2d} and~\ref{PiK1d} show that the scale  $T_{eff}$
is similar to a temperature in that
$\Pi^K$ is proportional to $\Omega$ for $\Omega>T_{eff}$ but  is proportional to $T_{eff}$ for $\Omega<T_{eff}$.
An electric field induced effective temperature was 
previously  identified in~\cite{Mitra06}, where the current flow across the layer 
controlled the decoherence, and in ~\cite{Green06} where the equivalent
of $\tau_{sc}$ was argued to arise from electron-magnon scattering which itself depended
on $T_{eff}$, leading to a nonlinear relation between $T_{eff}$ and $E$.

We now 
present a renormalization group treatment of Eq.~\ref{SK2}. 
The theory involves a momentum cutoff $\Lambda \sim k_F$
and we define the energy  unit to be $v_F\Lambda$ which we assume to be $>>T_{eff}$.
We also assume 
$v_D<v_F$ and $u,\delta<<1$.
We  
integrate out momenta in a small range $\Lambda$ to
$\Lambda e^{-l}$ 
and simultaneously integrate out all frequencies from 
$-\infty$ to $\infty$ (this procedure makes the preservation of  causality
manifest  \cite{Mitra06}). We  then
rescale momentum $q \rightarrow q e^{-l}$, the frequency $\omega \rightarrow \omega e^{-zl}$,
and the fields $m_{i} \rightarrow m_i e^{\alpha l}$, so as to keep the second and
third terms in Eq.~\ref{piReff} invariant, 
implying  dynamical exponent $z=2$. We have 
\begin{eqnarray}
\frac{dT_{eff}}{dl} &=& 2 T_{eff} \\
\frac{d\delta}{dl} &=& 2\delta +C_1 u_1 
\\
\frac{d v_D}{dl}& =& v_D \label{vDeq}\\
\frac{d u_i}{dl} &=& \left[2-d\right] u_i +  {\cal O}(u_k u_l) \label{ueq}
\end{eqnarray}
Here $C_1=lim_{v_F\Lambda/T_{eff}\rightarrow \infty}
\int_{-\infty}^{\infty}\frac{d\omega}{2\pi}
\frac{U \Pi_K(\omega, T_{eff})}
{U\Pi^R(q=\Lambda,\omega)U\Pi^A(q=\Lambda,\omega)}$.

Solving in dimension $d=2$ gives~\cite{Millis93,Mitra06} $T(l)=T_{eff}^0e^{2l}$, $v_D(l)=v_D^0e^l$,
$\delta(l)=r_0e^{2l}$ and $u_1(l)=2\pi^2/(9l)$ after initial transients have decayed.
Here the 
superscript $0$ denotes initial values and $r_0$ is related
to $\delta_0$ and $u$ as described in~\cite{Millis93}. The most interesting regime 
is the quantum critical/renormalized classical one occurring for $T^0_{eff}>|r_0|$. In this
case, at the scale $ l_T\approx ln \sqrt{1/T_{eff}^0}  $ a crossover to classical 
scaling occurs. Eq.~\ref{vDeq} shows that the drift term remains small ($v_D(l_T)\Lambda<<T(l_T)\sim 1$)
so that up to corrections of order $\sqrt{T_{eff}^0/E_F}$ the drift does not affect
the scaling.  The treatment of the crossover to the classical regime follows the procedure
given in~\cite{Mitra06} leading to 
a long wavelength  theory  described by the Halperin-Hohenberg model 
A dynamics, but with an extra
$\vec{j}\cdot{\vec{\nabla}}$ 
coupling to the order-parameter which represents current induced drift. Restoring physical units,
\begin{eqnarray}
&&\tau_{sc}\left( \frac{\partial}{\partial t} - {\vec v}_D\cdot
{\vec \nabla}\right)m_{cl}(x,t)
\nonumber \\
&&
=
- \left[r - A \nabla^2 +  \frac{2\pi^2m_{cl}^2}{9 ln \sqrt{1/T_{eff}^0}} \right]m_{cl} + \xi
\label{leq1} 
\end{eqnarray}
where 
\begin{equation}
\langle \xi(x,t) \xi(x^{\prime},t^{\prime}) \rangle =  \left(\frac{2\tau_{sc}T_{eff}}{\pi}\right) 
\delta(t-t^{\prime})\delta(x - x^{\prime})
\label{noisedef}
\end{equation}
The delta-correlated nature of the noise means  that
the low energy theory is Galilean invariant so that the drift may be eliminated
by the transformation $x \rightarrow x-v_Dt$. 

We now turn to the  case of a Heisenberg magnet 
which may be  treated along very similar lines except that we require a vector 
Hubbard-Stratonovich field \cite{Schulz90} which we denote by  ${\vec \Delta}$.
The physics is very similar to the Ising case except for fluctuations about the ordered
state. We concentrate on this in what follows. 
Expanding around $\vec{\Delta}_{cl} = \langle m_{cl} \rangle \hat{z},\vec{\Delta}_q=0 $ where
$\langle m_{cl} \rangle$ is given by Eq~\ref{saddle}, one obtains 
the Landau-Lifshitz-Gilbert 
equation of motion for the transverse spin wave fluctuations: 
\begin{equation}
\left(A{\bf 1}+iB\tau_y\right)
\begin{pmatrix} \Delta^x_{cl} \\ \Delta^y_{cl}
\end{pmatrix} =\xi \label{LLG}
\end{equation}
with $A=\alpha_{xx}\left[\frac{\partial}{\partial t} - \vec{v}_{Dxx}\cdot\vec{\nabla} \right] 
- C \nabla^2$ and $B=\alpha_{xy}\left[\frac{\partial}{\partial t} - \vec{v}_{Dxy}\cdot\vec{\nabla}\right] - D \nabla^2$
where $\xi$ is the white noise defined in Eq.~\ref{noisedef}.
The spin-wave correlation function obtained from Eq.~\ref{LLG} 
may be computed by standard methods and is found to evolve in time
as $e^{-F |t_1 - t_2|} $ with 
\begin{eqnarray}
\sl{Re}\left[F\right] = \frac{\left(\alpha_{xx}C  + \alpha_{xy} D \right)q^2\ + 
\alpha_{xx}\alpha_{xy}\left(\vec{v}_{D xx}-\vec{v}_{Dxy}\right)\cdot\vec{q} }{\alpha_{xx}^2 + \alpha_{xy}^2} 
\end{eqnarray}
showing that a dynamical instability signaled by $F^R < 0$ occurs 
if even an infinitesimally small current is applied to a magnet of  Heisenberg symmetry,
provided 
${\vec v}_{Dxx} \neq {\vec v}_{Dxy}$. 
For $v_{Dxx} = v_{Dxy}$, 
the ${\vec v}_D \cdot {\vec q}$ term may be eliminated via a Galilean transformation
and the current does
not drive a spin-wave instability. $v_{Dxx} - v_{Dxy}$ is a non-universal quantity.
In 
the model defined by Eqs \ref{hsys},\ref{hcoup} 
$v_{Dxx}$ differs from $v_{Dxy}$ only in the presence
of energy or momentum dependent scattering rates.

A derivation from microscopics of the current induced drift within a linear response treatment was presented in~\cite{Macd07,Tatara07}
for models with impurity scattering. Our treatment here differs in two ways, one is that we consider
a clean system where the dominant scattering mechanism is via an inelastic coupling to an external
reservoir. Secondly we go beyond linear response, which allows us to capture the effect of current
induced noise. Ref~\cite{Zhang05} showed that in the absence of noise Eq \ref{LLG}
has chaotic dynamics in the instability regime; the modifications of their results in the presence
of noise and critical fluctuations is an interesting open question. 

In summary
we have derived from a fundamental theory the effect of an in-plane current
on a magnetic quantum critical point. 
The current flow causes the critical fluctuations to drift 
with a velocity related to the spin current; however for Ising symmetry we find that in the regime
where critical fluctuations are important the theory is effectively Galilean-invariant so the
drift may be eliminated by an appropriate transformation.  In models with rotational invariance
the current leads to a spin wave instability whose critical behavior might be interesting
to study further; however
in the models we have studied the symmetry breaking effects are numerically small so the main
effect of the departure from equilibrium is an effective temperature, proportional to the voltage drop
over a distance of order of the mean free path. Extension of these results to the superconducting
case where the direct coupling between the order-parameter and the current can give rise to other nonequilibrium 
effects besides noise is currently in progress.

{\it Acknowledgements:}
The authors gratefully acknowledge useful discussions with Igor Aleiner. 
This work was supported by NSF-DMR-0705584.

\end{document}